# Quantifying Social Network Dynamics

Radosław Michalski, Piotr Bródka, Przemysław Kazienko, and Krzysztof Juszczyszyn

Institute of Informatics, Wrocław University of Technology, Wrocław, Poland
radoslaw.michalski@pwr.wroc.pl, piotr.brodka@pwr.wroc.pl, kazienko@pwr.wroc.pl, krzysztof@pwr.wroc.pl

*Abstract*—The dynamic character of most social networks requires to model evolution of networks in order to enable complex analysis of theirs dynamics. The following paper focuses on the definition of differences between network snapshots by means of Graph Differential Tuple. These differences enable to calculate the diverse distance measures as well as to investigate the speed of changes. Four separate measures are suggested in the paper with experimental study on real social network data.

*Keywords-social network changes, graph differential tuple, dynamics of the social network, SNA, graph edit distance*

## I. INTRODUCTION

Social networks and their analysis have become an extensively exploited domain of research. Usually, a social network is modelled as a graph, in which nodes correspond to social entities (people, group of individuals) while edges reflect relationships between those social entities. However, real social networks, extracted from data about user activities [7], have the dynamic nature. They evolve over time, and for that reason, there is a great need to model, analyse and measure the evolution of social network.

This work proposes a set of measures which values are capable to model the evolutionary patterns of the social network by measuring similarity between graphs. Those measures are evaluated as well to compare each other regarding the information scope they present about global changes of the social network. Proposed technique is also capable to analyse multi-layered social networks [7], [8]. This work extends and evaluates the concept presented in [24].

## II. RELATED WORK

When investigating the topological properties and structure of complex networks it is required to face complexity related problems. Large complex networks require significant computing overhead, for tasks like evaluating the centrality measurements, finding cliques, etc., what, generally, is a well-known fact. However, the technology-based social networks introduce new opportunities and approaches to solving the known problems of network analysis [11], [20], stemming from the idea of local topology analysis and partial problem solving.

This kind of networks (web communities, email social networks, user networks and so on) show important property which has a significant impact on the analysis – the existence of a link (connection) is typically a result of a series of discrete events associated with everyday human activity which have certain distribution in time. Human activity datasets are typically associated with the probability of interevent times (periods between the events, like sending an email) may be expressed as: $P(t) \approx t^{-\alpha}$ where typical values of $\alpha$ are from (1.5, 2.5), which result was confirmed for various communication technologies (email, phone calls, even for the exchange of paper letters etc.) [9]. Such distribution results in a series of consecutive events ("activity bursts") divided by longer periods of inactivity [1].

This phenomena is very important in the context of standard approach in dynamic network analysis, where network data are divided into time windows which are used to build series of time networks on the basis of data from given periods. Then the standard methods of social network analysis are applied to the consecutive networks which allows to observe how the chosen network characteristics (like centralities, network diameter, density etc.) change over time and possibly discover the underlying evolutionary patterns of the network [9], [14].

However, the abovementioned behaviour of the users (long inactivity periods mixed with the activity bursts) causes, in most cases, huge changes of any measure computed for neighbouring time windows. when we change the time windows. In result we observe a trade-off between choosing short windows which lead to chaotic changes of network measures, and long windows which offer stable results ignoring the network dynamics [2], [10].

In order to solve the problem, a number of methods designed to predict the changes in the structure of dynamic networks were proposed, some of them may be applied to the general link prediction problem [13], while the other address the periodicty of network changes, often observed in networks created from datasets reflecting user activity [12], [16].

In this paper, however, we are analysing networks from slightly different perspective – not a single link but the entire network viewed as a graph. Our proposal is to directly measure the differences between networks emerging in consecutive time windows. In this context our solution is a case of graph matching and graph similarity problems.

The general introduction and terminology for graph matching are presented in [5]. Additionally, graph matching methods were extensively discussed in [22], where structural similarity of local neighbourhoods was used to derive pairwise similarity scores between graph elements, and [21], [3], where discussion on basic notions of graph structural similarity was presented. Some work regarding modelling network dynamics, but mostly related to regenerate missing information was done in [23], however basing on exponential random graph model.

Comparing large graphs may be useful for integration and finding similarities between network layers. Algorithms for approximate matching of large graphs are proposed in [19]. A fast (and general enough to be applicable in large networks) method for attributed relational graphs (i.e. graphs with labels and semantics) is described in [4] along with algorithm definition and evaluation. An alternative approach to graph similarity for labelled, directed graphs, inspired by the simulation on labelled transition systems is reported in [18]. Functional graph similarity measure based on data fusion of the isomorphic and nonisomorphic subgraphs was proposed in [15]. A broad spectrum of graph similarity methods is used in image, video and pattern recognition [17]. The similarities in local topology of temporal graphs were analysed in [6].

### III. THE DIFFERENCE OF GRAPHS GRAPH DIFFERENTIAL TUPLE

#### A. General Concept

Two graphs can differ in many ways. There can be different vertices, different edges and – for edges between the same vertices – different weights. In this concept we want to define the difference of graphs – basic operations needed to be done in order to transform one graph into the other one – see Figure 1.

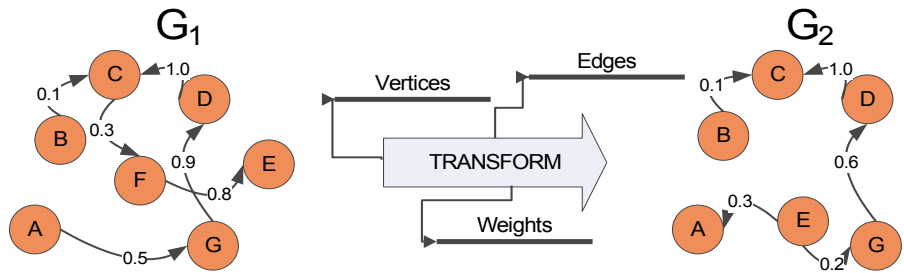

Figure 1. General Concept of transforming one graph to another

Two graphs can be defined as:

$G_1 = \langle V_1, E_1 \rangle$ and $G_2 = \langle V_2, E_2 \rangle$, where:
$V_1$ – set of vertices in graph $G_1$,
$V_2$ – set of vertices in graph $G_2$,
$E_1 = \{\langle x,y \rangle : x, y \in V_1\}$ – edges in $G_1$,
$E_2 = \{\langle x,y \rangle : x, y \in V_2\}$ – edges in $G_2$,
$w(x, y) \to [0,1]$ – weight of the edge between $x$ and $y$.

Using the above definition, the difference between graphs can be introduced as a set of different vertices, different edges and different weights.

#### B. Graph Differential Tuple

In order to present graph difference in a consistent, detailed shape, the Graph Differential Tuple can be defined as follows:

$G_1 G_2 = \langle V^+, V^-, E^+, E^-, E^\Delta \rangle$, where:
$V^+ = \{x : x \notin V_1 \land x \in V_2\}$ – set of added vertices
$V^- = \{x : x \in V_1 \land x \notin V_2\}$ – set of removed vertices
$E^+ = \{\langle x,y \rangle : \langle x,y \rangle \notin E_1 \land \langle x,y \rangle \in E_2\}$ – set of added edges
$E^- = \{\langle x,y \rangle : \langle x,y \rangle \in E_1 \land \langle x,y \rangle \notin E_2\}$ – set of removed edges
$E^\Delta$ – the set of modified weight tuples $\langle\langle a, b\rangle, w_m(a, b)\rangle$

#### C. Case Study

Using Graph Differential Tuple, the evolution of a social network can be presented. Having e.g. two time windows of the same social network, this approach would reveal differences and show how this social network is changing in time, answering following set of questions:

- Who is new in the network?
- Who is no longer in the network?
- Where are new connections in the network?
- Which connections have disappeared?
- Which connections are now stronger/weaker?

TABLE I. WEIGHT MATRIX FOR EDGES OF GRAPHS $G_1$ AND $G_2$

| Edge | Weight | |
|------|--------|--|
|      | Graph $G_1$ | Graph $G_2$ |
| ⟨A,B⟩ | 0.3 | 0.3 |
| ⟨A,G⟩ | -   | 0.3 |
| ⟨B,A⟩ | 0.5 | 0.5 |
| ⟨B,C⟩ | 0.8 | 0.8 |
| ⟨C,D⟩ | 1.0 | -   |
| ⟨C,E⟩ | 0.7 | 0.3 |
| ⟨D,C⟩ | 0.9 | -   |
| ⟨D,E⟩ | 0.2 | -   |
| ⟨E,C⟩ | -   | 0.1 |
| ⟨E,D⟩ | 0.1 | -   |
| ⟨F,B⟩ | 0.6 | 0.9 |
| ⟨F,E⟩ | 0.4 | -   |
| ⟨G,A⟩ | -   | 0.4 |

Let $G_1$ and $G_2$ be defined as follows:
$G_1 = \langle V_1, E_1 \rangle$
$V_1 = \{A, B, C, D, E, F\}$
$E_1 = \{\langle A,B\rangle, \langle B,A\rangle, \langle B,C\rangle, \langle C,D\rangle, \langle D,C\rangle, \langle D,E\rangle, \langle E,D\rangle, \langle C,E\rangle, \langle F,B\rangle, \langle F,E\rangle\}$
$G_2 = \langle V_2, E_2 \rangle$
$V_2 = \{A, B, C, E, F, G\}$
$E_2 = \{\langle A,B\rangle, \langle B,A\rangle, \langle B,C\rangle, \langle C,E\rangle, \langle E,C\rangle, \langle F,B\rangle, \langle G,A\rangle, \langle A,G\rangle\}$

Then the Graph Differential Tuple will look as follows:
$G_1G_2 = \langle V^+, V^-, E^+, E^-, E^\Delta \rangle$, where
$V^+ = \{G\}$
$V^- = \{D\}$
$E^+ = \{\langle E, C \rangle, \langle G, A \rangle, \langle A, G \rangle\}$
$E^- = \{\langle C, D \rangle, \langle D, C \rangle, \langle D, E \rangle, \langle E, D \rangle, \langle F, E \rangle\}$
$E^\Delta = \{\langle \langle C, E \rangle, -0.4 \rangle, \langle \langle F, B \rangle, 0.3 \rangle\}$

## IV. THE MEASURES OF DISTANCE BETWEEN GRAPHS

Based on the distance vector $G_1G_2 = \langle V^+, V^-, E^+, E^-, E^\Delta \rangle$ described in previous section, a few distance measures, which presents the distance between two graphs in numbers can be introduced.

### A. The Sum

The sum measure is the simplest measure possible. It is represented by weighted sum of all sets from $G_1G_2$ vector, i.e.:

$$d^s(G_1, G_2) = \alpha^+|V^+| + \alpha^-|V^-| + \beta^+|E^+| + \beta^-|E^-| + \gamma|E^\Delta| \quad (1)$$

where: $\alpha^+, \alpha^-, \beta^+, \beta^-, \gamma \in \{0;1\}$ are the coefficients which reflect the importance of each vector element.

### B. The Normalized Sum

The second measure is based on the first one but it is normalised by the number of the nodes and edges from both graphs. It returns value from range [0;1], where 0 means that two graphs are identical, and 1 that graphs are completely different. It is defined as follows:

$$d^n(G_1, G_2) = \frac{\alpha^+|V^+| + \alpha^-|V^-| + \beta^+|E^+| + \beta^-|E^-| + \gamma|E^\Delta|}{|V_1| + |V_2| + |E_1| + |E_2|} \quad (2)$$

### C. The Relative Sum

The relative sum informs how the graphs do differ, but relatively to the first graph:

$$d^{fn}(G_1, G_2) = \frac{\alpha^+|V^+| + \alpha^-|V^-| + \beta^+|E^+| + \beta^-|E^-| + \gamma|E^\Delta|}{|V_1| + |E_1|} \quad (3)$$

### D. Based on Edge Modification

The last measure is built on $E^\Delta$, i.e., edges modifications and computed as follows:

$$d^{mw}(G_1, G_2) = \begin{cases} \dfrac{\sum_{a,b \in N_1 \cap N_2} mw(a,b)}{|E_1 \cap E_2|} & \text{when } E_1 \cap E_2 \neq \phi \\ 0 & \text{in the other case} \end{cases} \quad (4)$$

where: $mw(a,b) = |w_2(a,b) - w_1(a,b)|$

## V. EXPERIMENTAL STUDIES

### A. Data Sets

The experiments were conducted on the data gathered from Wroclaw University of Technology email communication (among staff members). The whole data set was collected within period of February 2006 – October 2007 and consist of 5,845 nodes and 149,344 edges.

The data was split into two data sets. The first one consisting of forty 30-days slot social networks. The social network slots are overlapping with the 15-day overlap, i.e. the first social network slot begins on the $1^{st}$ day and ends on the $30^{th}$ day, the second one starts on the $16^{th}$ day and lasts till the $45^{th}$ day and so on. The second data set consist of twenty 30-day slot social networks but this time the social network slots are not overlapping, i.e., the first slot begins on the 1st day and ends on the $30^{th}$ day, the second one lasts from the $31^{st}$ day until the $60^{th}$ day and so on.

The weight $w_i(x,y)$ of the edge from node $x$ to $y$ in the $i^{th}$ social network (in the $i^{th}$ time slot) was calculated separately for each edge in each social network, as follows:

$$w_i(x, y) = \frac{N_i(x, y)}{N_i(x)} \quad (5)$$

where $N_i(x,y)$ represents amount of e-mails sent by $x$ to $y$ and $N_i(x)$ represents total amount of e-mails sent by $x$.

### B. Parameters

To calculate the first three measures introduced in Section 4 (Eq. 1, 2, and 3), a matrix of distinct combinations for $\alpha^+, \alpha^-, \beta^+, \beta^-$ and $\gamma$ parameters was defined. The values of the parameters were given only from the set $\{0,1\}$ to represent only the border cases – the element of each sum could be either taken into account with the value of 1 or discarded at all. It is due to the fact that we are more interested in using or discarding particular elements of the Graph Differential Tuple rather than respecting their influence in a more smooth way. All the combinations analysed are presented in Table II, where columns represent combination indices and rows - parameters.

TABLE II. COMBINATIONS OF THE PARAMETERS (SEE EQ. 1, 2, 3)

|    | $\alpha^+$ | $\alpha^-$ | $\beta^+$ | $\beta^-$ | $\gamma$ |    | $\alpha^+$ | $\alpha^-$ | $\beta^+$ | $\beta^-$ | $\gamma$ |
|----|---|---|---|---|---|----|---|---|---|---|---|
| 1  | 1 | 0 | 0 | 0 | 0 | 17 | 0 | 1 | 0 | 1 | 1 |
| 2  | 0 | 1 | 0 | 0 | 0 | 18 | 0 | 1 | 1 | 0 | 1 |
| 3  | 0 | 0 | 1 | 0 | 0 | 19 | 0 | 1 | 1 | 1 | 0 |
| 4  | 0 | 0 | 0 | 1 | 0 | 20 | 1 | 0 | 1 | 1 | 0 |
| 5  | 0 | 0 | 0 | 0 | 1 | 21 | 1 | 0 | 0 | 1 | 1 |
| 6  | 1 | 1 | 0 | 0 | 0 | 22 | 1 | 0 | 1 | 0 | 1 |
| 7  | 1 | 0 | 1 | 0 | 0 | 23 | 1 | 1 | 0 | 1 | 0 |
| 8  | 1 | 0 | 0 | 1 | 0 | 24 | 1 | 1 | 0 | 0 | 1 |
| 9  | 1 | 0 | 0 | 0 | 1 | 25 | 1 | 1 | 1 | 0 | 0 |
| 10 | 0 | 1 | 1 | 0 | 0 | 26 | 1 | 1 | 1 | 1 | 0 |
| 11 | 0 | 1 | 0 | 1 | 0 | 27 | 0 | 1 | 1 | 1 | 1 |
| 12 | 0 | 1 | 0 | 0 | 1 | 28 | 1 | 0 | 1 | 1 | 1 |
| 13 | 0 | 0 | 1 | 1 | 0 | 29 | 1 | 1 | 0 | 1 | 1 |
| 14 | 0 | 0 | 1 | 0 | 1 | 30 | 1 | 1 | 1 | 0 | 1 |
| 15 | 0 | 0 | 0 | 1 | 1 | 31 | 1 | 1 | 1 | 1 | 1 |
| 16 | 0 | 0 | 1 | 1 | 1 |    |   |   |   |   |   |

*C. Results*

For both data sets, all the measures, introduced in Section IV, were calculated - each social network was compared with the previous one, i.e. the second with the first, the third with the second, etc.

Their values normalised by the maximum value of each measure, for selected two parameter combinations, i.e. 7 (added nodes and edges) and 31 (every component respected). They are presented in Fig. 2a-2d.

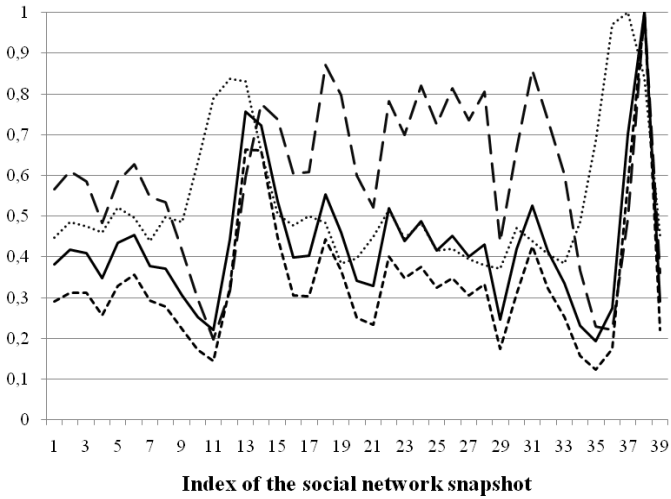

Figure 2a. Comparison of measures - the first dataset, the 7$^{th}$ combination of parameters

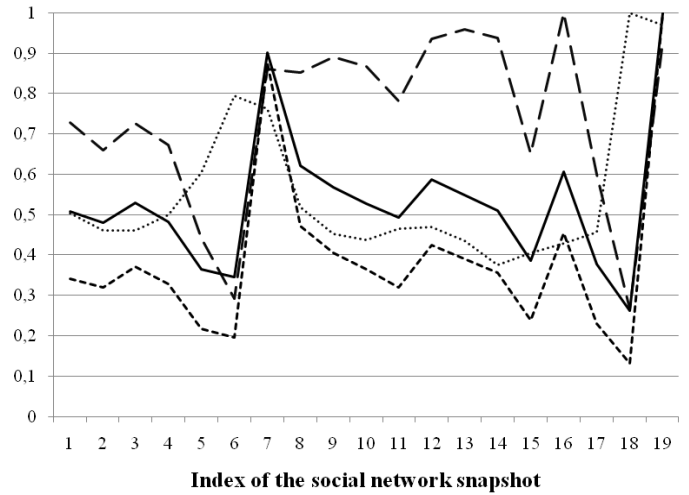

Figure 2b. Comparison of measures - the second dataset, the 7$^{th}$ combination of parameters

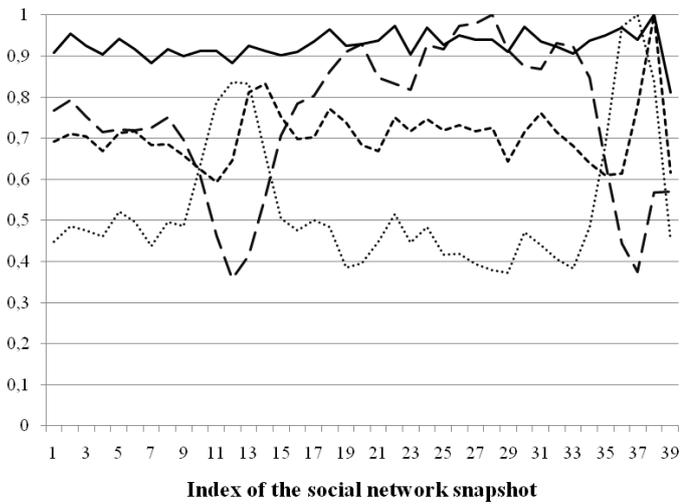

Figure 2c. Comparison of measures - the first dataset, the 31$^{st}$ combination of parameters

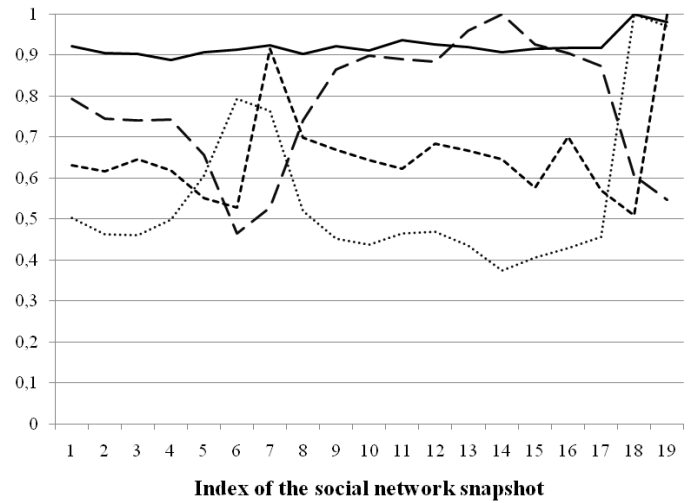

Figure 2d. Comparison of measures - the second dataset, the 31$^{th}$ combination of parameters

— — Sum    ——— Normalised Sum    ----- Relative Sum    ······· Edge Modification

In Figures 3a-3f differences between parameter-dependent measures' values normalised to the maximum value for both datasets according to four chosen parameter combinations: 7, 14, 26, 31 are presented.

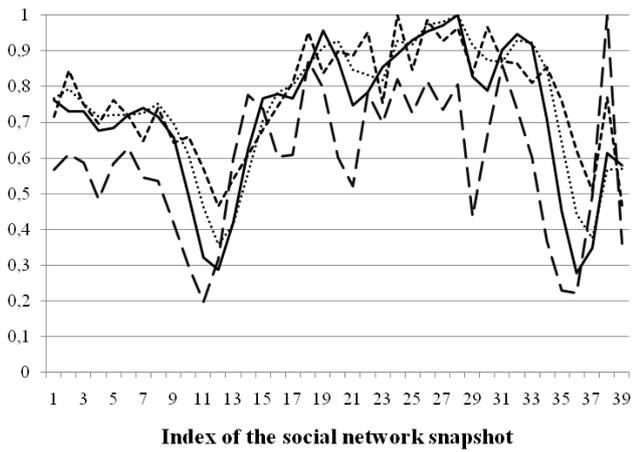

Figure 3a. Parameter-dependent measure values, the first dataset, the Sum

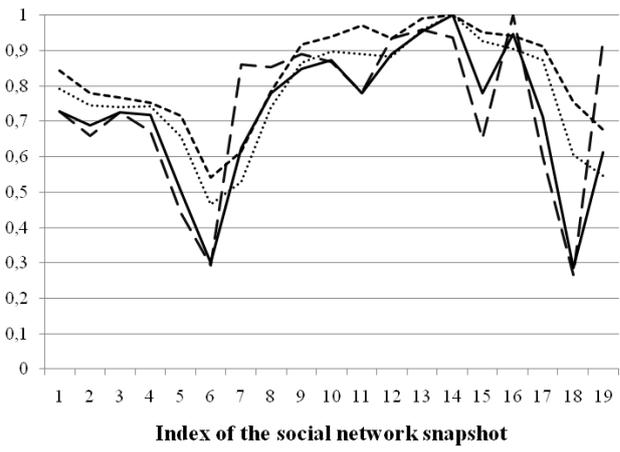

Figure 3b. Parameter-dependent measure values, the second dataset, the Sum

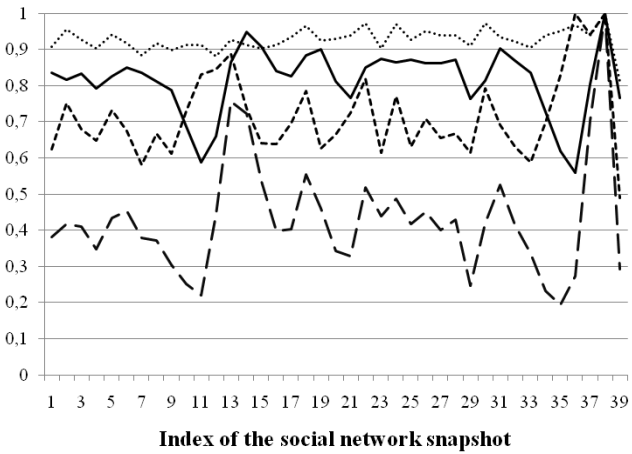

Figure 3c. Parameter-dependent measure values, the first dataset, the Normalised Sum

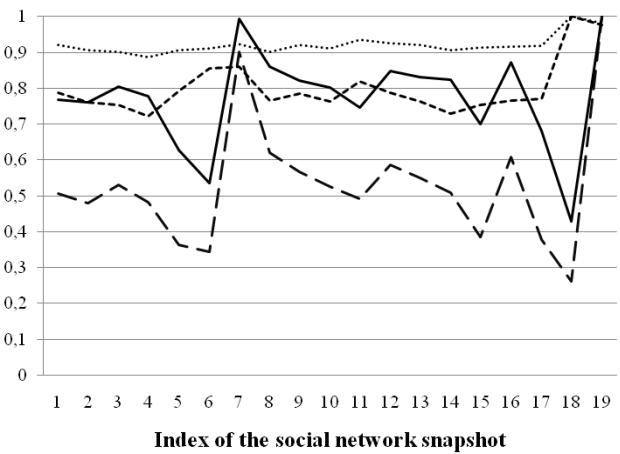

Figure 3d. Parameter-dependent measure values, the second dataset, the Normalised Sum

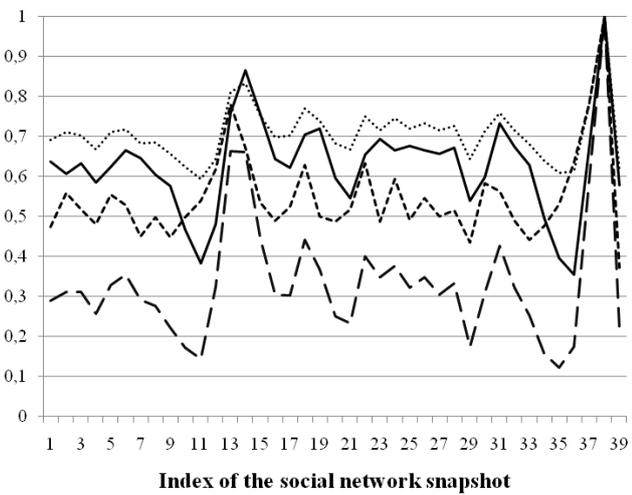

Figure 3e. Parameter-dependent measure values, the first dataset, the Relative Sum

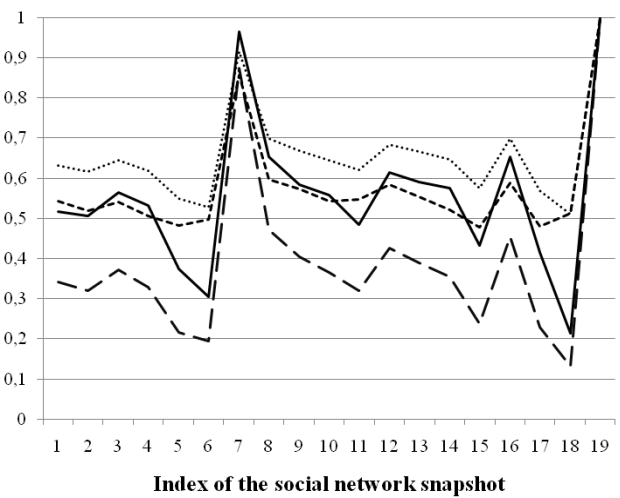

Figure 3f. Parameter-dependent measure values, the second dataset, the Relative Sum

— — 7th combination       ——— 14th combination       - - - - 26th combination       ········ 31st combination

## VI. Conclusions and Future Work

A new concept for network evolution modelling is presented in this paper. It is based on comparison of two graphs and its essential component is Graph Differential Tuple, which preserves changes between the second and the first graph. Having Graph Differential Tuple extracted, four distinct distance measures were proposed: simple sum, normalized sum, relative sum and edge modification sum. All of them describe the change (difference between two graphs) by means of simple numerical values. The first three contain parameters, which in fact allow to create a variety of measures by inclusion, exclusion and emphasising their different components. The reason behind introducing such a measures set is that they extend the well-known idea of graph edit distance in terms of emphasising particular aspects of the network evolution direction.

In the experimental studies the authors decided to calculate proposed measures using two datasets based on the same social network source – e-mail communication. The goal was to obtain the knowledge how well the introduced measures are modelling the social network evolution: are they similar to each other or do they differ completely? are they responding fast for changes or not? Results show that for the first three measures the combination of parameters strongly influences the similarity of values. It means that at this point of research it is rather recommended to selectively choose the parameters in proposed measures rather than to mix all type of changes (i.e. only to analyse additions, not additions and deletions altogether). The second experiment confirmed that the combination of parameters does not influences the overall view of network changes – they do behave similar. The last analysis shows that it is not necessary to use overlapping windows for the analysis of social network dynamics, when using proposed measures, because they do differ slightly. In that case the measures will be calculated faster because of smaller dataset, built using non-overlapping windows.

Future work will focus on development of new measures, effective algorithms of their computation for huge networks as well as on their application to new real data sets. Some more concern would be also given to the $E^\Delta$ set, which should provide more information about the level of change of edges.

## Acknowledgment

The work was partially supported by fellowship co-financed by the European Union within the European Social Fund, The Polish Ministry of Science and Higher Education, the research project 2010-13.